\documentclass[10pt,aps,pra,superscriptaddress,notitlepage,onecolumn]{revtex4-2}
\usepackage{hyperref}
\usepackage{amsmath}
\usepackage{graphicx}
\usepackage{color}
\usepackage{braket}
\usepackage{epstopdf} 
\usepackage{multirow}

\usepackage[paperwidth=210mm,paperheight=297mm,centering,hmargin=3cm,vmargin=2.5cm]{geometry}

\newcommand{\be}{\begin{equation}}
\newcommand{\ee}{\end{equation}}

\usepackage{bm}

\newcommand{\bs}{\boldsymbol}

\newcommand{\addcqt}{Centre for Quantum Technologies, National University of Singapore, 3 Science Drive 2, Singapore 117543}
\newcommand{\addtuc}{School of Electrical and Computer Engineering, Technical University of Crete, Chania, Greece 73100}
\newcommand{\addaq}{AngelQ Quantum Computing, 531A Upper Cross Street, \#04-95 Hong Lim Complex, Singapore 051531}

\begin{document}

\title{Topological data analysis and machine learning}

\author{Daniel Leykam}
%\email{daniel.leykam@gmail.com}
\affiliation{\addcqt}
\author{Dimitris G. Angelakis}
\affiliation{\addcqt}
\affiliation{\addtuc}
\affiliation{\addaq}

\date{\today}

\begin{abstract}
Topological data analysis refers to approaches for systematically and reliably computing abstract ``shapes'' of complex data sets. There are various applications of topological data analysis in life and data sciences, with growing interest among physicists. We present a concise review of applications of topological data analysis to physics and machine learning problems in physics including the unsupervised detection of phase transitions. We finish with a preview of anticipated directions for future research.
\end{abstract}

\maketitle

\section{Introduction}

Topological quantities are invariant under continuous deformations; an often-cited example is that a doughnut can be continuously transformed into coffee mug - both are topologically equivalent to a torus. The robustness of topological quantities to perturbations is inspiring physicists in many fields, including condensed matter, photonics, acoustics, and mechanical systems~\cite{RevModPhys.82.3045,RevModPhys.91.015006,Ma2019,Kim2020}. In all these areas topology has enabled the prediction and explanation of surprisingly robust physical effects.

Most famously, the extremely precise quantisation of the Hall conductivity observed in two-dimensional electronic systems since the 1980s was explained as a novel topological phase of matter, the quantum Hall phase~\cite{vonKlitzing2020}. In this and many other examples from physics, we deal with smooth deformations in some parameter space, such as the energy bands of solid state electronic systems.

Physics is however an outlier among fields of science in that idealised continuous models and functions can explain a wide variety of observed phenomena. Other fields do not have the luxury of continuity and have to make do out of sparse data and limited observations in high dimensional parameter spaces. Despite this very different setting, topological approaches remain powerful. 

A suite of computational topological techniques known as topological data analysis (TDA) has been developed over the past twenty years to systematically define and study the ``shape'' of complex discrete data in high dimensional spaces. TDA is attracting growing interest among physicists, particularly those working on topological materials or the application of machine learning techniques to physics~\cite{RevModPhys.91.045002,MEHTA20191,doi:10.1080/23746149.2020.1797528,doi:10.1080/23746149.2022.2046156,https://doi.org/10.48550/arxiv.2204.04198}. 

At this time we are aware of two existing reviews on TDA aimed at the physics audience. The first by Carlsson, one of the founders of the field, gave a broad survey of different techniques of TDA and their applications in various areas of science~\cite{Carlsson2020}. The second review, by Murugan and Robertson, provided a detailed pedagogical and physicist-friendly introduction to two important techniques, persistent homology and the Mapper algorithm, applying them to the example of an astronomical dataset~\cite{https://doi.org/10.48550/arxiv.1904.11044}.

Since publication of these two reviews there has been growing interest in applying TDA methods to physics, including the incorporation of TDA into physics-targeted machine learning, with applications including the unsupervised detection of phase transitions. Moreover, the field of TDA has continued to evolve with new generalisations and techniques being actively studied.

The aim of this article is to review cutting edge applications of TDA to physics. We will provide a gentle introduction to the basic techniques, survey how TDA shows promise for the detection of novel phases of matter, and speculate on what we believe to be important directions for future research, including opportunities offered by newer TDA methods such as zigzag persistence.

The structure of this article is as follows: Sec.~\ref{sec:tda} provides a brief introduction to TDA guided by the simple example of two-dimensional point clouds. Sec.~\ref{sec:physics_applications} discusses how TDA has been applied to identify order parameters and phase transitions in various physical systems. Sec.~\ref{sec:TDA_ML} covers recent studies that employ TDA to compute features of physical systems that are then incorporated into a larger machine learning pipeline. Sec.~\ref{sec:future} speculates on anticipated future directions and applications of TDA to physics, and vice versa. We conclude with Sec.~\ref{sec:conclusion}.

\section{Topological data analysis}
\label{sec:tda}

Admittedly TDA has a rather steep learning curve, since its foundation differs from the topology of continuous spaces to which physicists are more accustomed. However, after battling through the unfamiliar jargon and notation one can develop a powerful intuition for the subject. Our aim here is to give an equation-free sketch of the general approaches and terminology, while referring the motivated reader to more comprehensive and mathematically-rigorous reviews~\cite{Carlsson2020,https://doi.org/10.48550/arxiv.1904.11044,Carlsson2009,doi:10.1146/annurev-statistics-031017-100045,10.3389/frai.2021.681108}.

\subsection{From point clouds to persistence diagrams}

\begin{figure}

\includegraphics[width=\columnwidth]{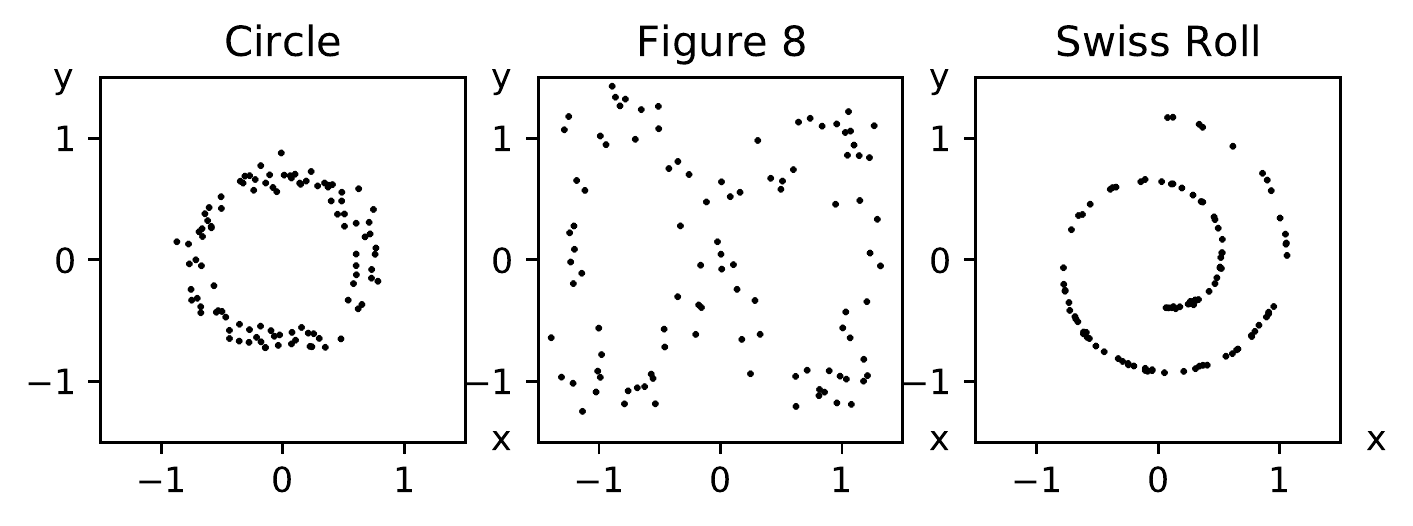}

\caption{Examples of noisy point clouds. Point clouds sampled from objects with differing shapes and even differing dimensionality may be difficult to distinguish using standard summary statistics such as the centre of mass and variance. In ``Circle'' and ``Figure 8'' the noise randomly perturbs the points in the ambient two-dimensional space. In ``Swiss Roll'' points are sampled from a one-dimensional interval before being embedded into the two-dimensional space $(x,y)$.}

\label{fig:clouds}

\end{figure}

As an instructive example let us consider the two-dimensional point clouds shown in Fig.~\ref{fig:clouds}. Each point may correspond to a distinct measurement of some object, e.g. the locations of photons arriving at a camera, or the positions of particles in a system. With our eyes we can clearly see that each cloud has a different shape: The points in the ``Circle'' and ``Figure 8'' clouds are distributed around one and two loops, respectively. On the other hand, the ``Swiss Roll'' corresponds to a noisy one-dimensional point cloud embedded into a higher (two-) dimensional space.

We would like to formalise these qualitative observations in a more systematic way, such that we are not reliant on directly plotting the data, which is an approach limited to two- or three-dimensional datasets. How can we quantify the obviously different shapes of these point clouds? Standard summary statistics such as the centre of mass or variance are clearly inadequate, since they are not invariant under shape-preserving translations or rescaling of the data.

Fortunately graph theory provides rigorous ways of quantifying intuitive shapes of discrete datasets including point clouds. The idea is to construct a graph by connecting pairs of points (vertices) that are sufficiently close together by edges, and then quantify shape by computing topological invariants of the graph, its Betti numbers $B_k$. The $k$th Betti number is the number of $k$-dimensional holes, e.g. the number of independent connected components (clusters $B_0$) or non-contractible loops (cycles $B_1$). In practice, evaluating graph invariants amounts to computing the ranks and null spaces of linear operators (matrices) acting on the graph's vertices and edges. In a nutshell, the computation of the shape of point cloud data can be reduced to simple linear algebra.

Higher-dimensional topological features can be similarly obtained by constructing generalisations of graphs known as simplicial complexes, which capture higher dimensional objects (faces, volumes, etc.) by triangulation. A $k$-simplex is a combination of $(k+1)$ vertices; edges are 1-simplices, triangular faces are 2-simplices, tetrahedral volumes are 3-simplices, and so on. A $k$-simplicial complex is collection of simplices with dimension of at most $k$.

Increasing $k$ complicates matters. First, since $k$-simplices are combinatorial objects the number of possible simplices grows rapidly with $k$, limiting practical calculations to low dimensional topological features. Second, there is no unique way to construct a simplicial complex given only pairwise distances between points and a cutoff scale; different methods may differ in their computational costs, stability properties, and ability to faithfully reproduce shapes of the underlying space from which the points are sampled~\cite{hausmann1995vietoris}.

\begin{figure}

\includegraphics[width=\columnwidth]{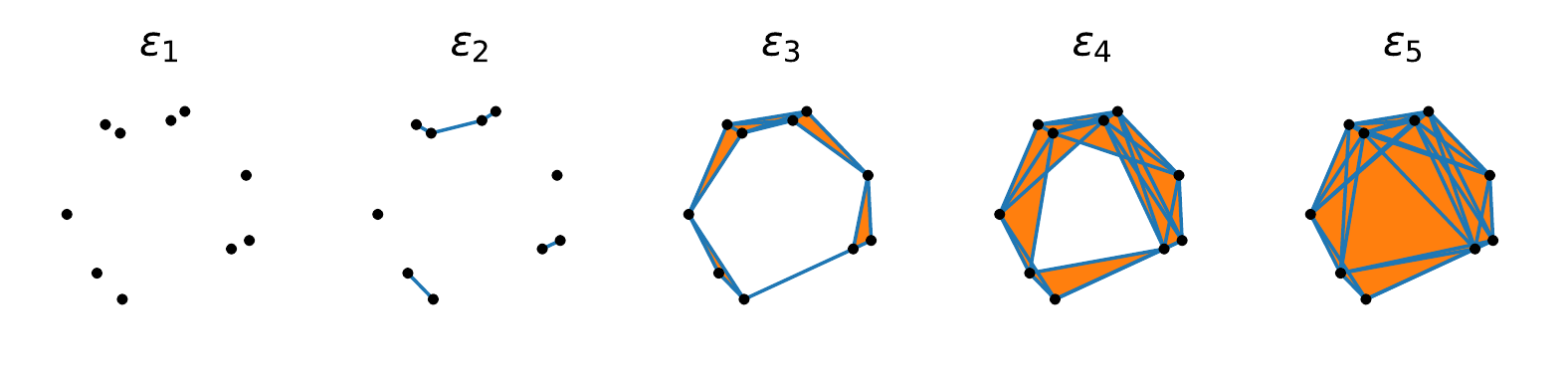}

\caption{Simplicial complexes constructed from a point cloud using different cutoff distances $\epsilon_i$, where blue lines and orange shaded areas denote edges and faces, respectively. For small cutoff distances all points are disconnected, forming a trivial simplicial complex with no edges ($\epsilon_1$). As the cutoff is increased nearby vertices start to become connected by edges ($\epsilon_2$). Increasing the cutoff further, triplets of points become connected, forming faces. In $\epsilon_3$ and $\epsilon_4$ the simplicial complex has a single connected component hosting a non-trivial cycle. For sufficiently large cutoff distances the cycle is destroyed by the addition of faces covering the entire interior of the point cloud ($\epsilon_5$).}

\label{fig:filtration}

\end{figure}

There is one big elephant in the room we must address: what do we mean by ``sufficiently close'' when connecting vertices to form the graph or simplicial complex? How do we determine which pairs of vertices to link by an edge and which pairs to leave disconnected? The number of cycles and clusters will be sensitive to the choice of cutoff distance and even possibly the addition or removal of a single edge, as illustrated in Fig.~\ref{fig:filtration}. This seems like a big problem making the approach lack robustness to noise and other perturbations.

The neat solution to the scale-dependence of graph invariants obtained from point clouds is to compute the shape of the graph over an entire range of scales known as a filtration, i.e. study its topology as a function of the cutoff length scale~\cite{Robins1999}. Topological features (e.g. clusters, cycles) persisting over a wide range of scales are more robust and should provide a meaningful characterisation of the overall shape of the data. On the other hand, features sensitive to small changes in scale or the addition or removal of a few edges can be attributed to noise and discarded if necessary. By studying the persistence of topological features we will be able to distinguish robust features from noise.

\begin{figure}[b]

\includegraphics[width=\columnwidth]{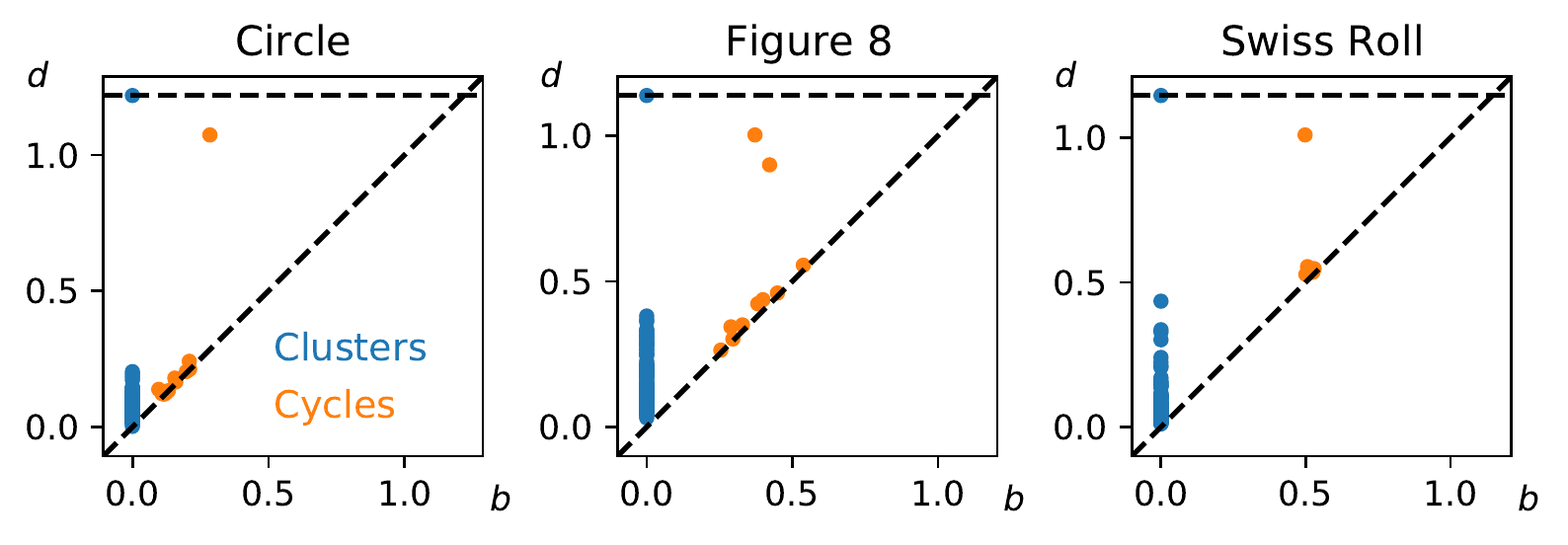}

\caption{Persistence diagrams of the two-dimensional point clouds shown in Fig.~\ref{fig:clouds} computed using the Vietoris-Rips complex~\cite{hausmann1995vietoris}. Each point represents a distinct topological feature. Horizontal and vertical axes denote the length scales at which each feature is created ($b$; birth) and destroyed ($d$; death) respectively. Points that are further from the diagonal dashed line therefore persist over a larger range of scales and are said to have a longer ``lifetime'' $l = d-b$. Since features must be created before they are destroyed, no points lie below the diagonal. At sufficiently large spatial scales all points become connected to form a single connected graph, corresponding to a single cluster with an infinite lifetime. Typically the infinite lifetime cluster is either discarded or plotted at a finite $d$ and distinguished using a horizontal dashed line.}

\label{fig:pds}

\end{figure}

Persistence diagrams are one stable way to represent scale-dependent topological features of a dataset~\cite{Cohen-Steiner2007}. Fig.~\ref{fig:pds} shows persistence diagrams computed for each of the point clouds in Fig.~\ref{fig:clouds}. The most persistent topological features not only allow us to infer the overall shape of the data, but also gives information as to the geometry of the point cloud. For example, the birth scales of the long-lived cycles in the ``Circle'' and ``Figure 8'' clouds are related to the a maximum separation between neighbouring points comprising the cycle, while the death scale will be related to the cycle's diameter.

The attentive reader will notice that the persistence diagrams for the ``Circle'' and ``Swiss Roll'' clouds share the same long-lived features, despite their obviously-differing shapes. Closer inspection will, however, reveal noticeable differences in their short-lived features. For example, the cycles appearing in the ``Swiss Roll'' dataset all have similar birth scales, corresponding to the distance between the inner and outer part of the spiral and hinting at a one-dimensional embedding. This suggests that the differing shapes of these two point clouds may indeed be captured by inspecting their short-lived features; thus, persistent homology can also capture the local features (geometry) of the data.

\subsection{Comparing and computing persistence diagrams}
\label{sec:vectorisation}

While persistence diagrams provide a compact visual summary of the scale-dependent topological features of a single dataset, it is not immediately clear how we should go about comparing persistence diagrams computed for different datasets; they will generally differ in their number of features and level of noise, making it difficult to establish a common threshold between genuine features and noise-induced features.

These issues motivated the development of stable distance and similarity measures for persistence diagrams. Here stability means that a small change to one dataset results in, at most, a similarly small change to the similarity to other fixed persistence diagrams.

\begin{figure}

\includegraphics[width=\columnwidth]{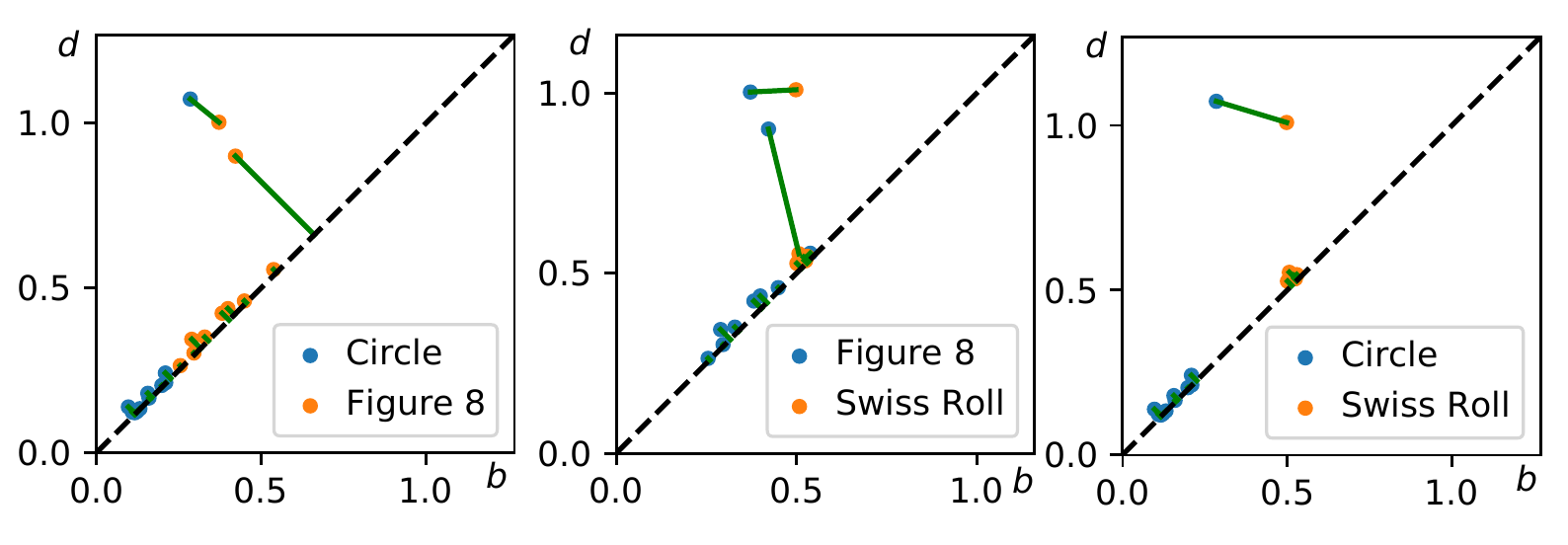}

\caption{Matching (green lines) of the one-dimensional cycles of the Circle, Figure 8, and Swiss Roll point clouds used to compute the Wasserstein distance, which corresponds to the total length of the green lines.}

\label{fig:matchings}

\end{figure}

One example of a stable distance measure is the Wasserstein distance, which is the smallest distance the points in a pair of persistence diagrams must be moved in order to transform one diagram into the other. Unpaired features (i.e. if one diagram has more features) are moved to the diagonal. For example Fig.~\ref{fig:matchings} shows the matching between the one-dimensional cycles of the Circle, Figure 8, and Swiss Roll point clouds. Since all features contribute to the Wasserstein distance, even the noise-induced ones close to the diagonal, it can be less sensitive to changes in the most persistent features. Another popular choice of distance measure is the bottleneck distance, which is the largest deformation of a pair of features required to convert one diagram to another (i.e. the Wasserstein distance under the $p=\infty$ norm). The bottleneck distance is thus independent of the short-lived features near the diagonal.

Alternative approaches for characterising and comparing the information contained in the persistence diagrams employ vectorisation: the variable length information encoded in the $(b,d)$ pairs of the persistence diagram are mapped to a vector or vectors in a fixed-dimensional space; different persistence diagrams can then be studied using more familiar tools such as vector inner products. For example, one might compute a set of summary statistics such the entropy or moments of the feature lifetimes $l = |d-b|$~\cite{CHINTAKUNTA2015391,10.1007/978-3-319-29228-1_11,ATIENZA2020107509}, assuming they are relevant to the task at hand. 

Often the relevant features are unknown a priori and it is preferable to compute a high-dimensional vectorisation to minimise the loss of relevant information. For example, the persistence landscape provides a stable and invertible (i.e. information-preserving) vectorisation of persistence diagrams~\cite{JMLR:v16:bubenik15a,BUBENIK201791}.

Using a distance measure or vectorisation allows one to combine persistent homology with powerful machine learning techniques such as artificial neural networks or clustering algorithms to compare topological features of different datasets and perform tasks including shape-based identification and classification of different point clouds, which will be explained further in Sec.~\ref{sec:TDA_ML}. However, one important consideration in applying vectorisation or distance measures is that they can introduce additional hyper-parameters that may affect the sensitivity to different topological features of the data. 

There are a variety of software libraries for computing persistence diagrams, their vectorisation, and distance measures~\cite{10.1007/978-3-662-44199-2_28,ctralie2018ripser,giotto-tda,Bauer2021}, surveyed in Ref.~\cite{Otter2017}. Crucial for applications, persistence diagrams can be efficiently computed given a filtration by building up the simplicial complex one element at a time, detecting any changes to the topological features at each step. This yields not only the feature birth and death scales, but also their representations, e.g. edges comprising a cycle. Nevertheless, due to the combinatorial nature of simplicial complexes the computational requirements grow rapidly with the feature dimension $k$, with most practical applications limited to $k \leq 2$. 

To compute a persistence diagram the end-user must provide at a minimum either the data points or a distance matrix encoding pairwise distances between points. One can also consider custom filtrations. For example, when dealing with image data one can use the greyscale pixel values as a filtration parameter, constructing a simplicial complex out of pixels less than (or exceeding) a given threshold~\cite{5613465,Robins2011}. The resulting sublevel (superlevel) set filtration summarises the critical points of an image, i.e. its local minima, maxima, and saddle points, as well as their higher-dimensional generalisations. 

\subsection{Other approaches and recent developments}

The above discussion of persistent homology has been limited to the simplest case of simplicial complexes constructed from two-dimensional point clouds. There are a variety of related techniques for studying complex datasets by reducing them to families of graphs or simplicial complexes which we only mention briefly here due to space constraints.

The Mapper algorithm reduces point clouds to simpler low-dimensional graphs by performing clustering on overlapping subsets of the data~\cite{10.2312:SPBG:SPBG07:091-100,doi:10.1063/1.3103496}. Local anomalies such as intersections and cusps can be similarly detected by comparing the persistent homology of different subsets of the data~\cite{doi:10.1073/pnas.2001741117}.

Standard persistent homology constructs filtrations as a sequence of nested simplicial complexes; as the filtration parameter (e.g. cutoff distance) is increased edges and higher-dimensional simplices are added to the complex and never removed. In certain situations, e.g. when studying temporal network dynamics, simplices can be both added and removed as a control parameter is varied. Zigzag persistence is a technique that enables the identification of significant topological features in this case~\cite{Carlsson2010}.

Another important problem is to compute persistent topological features as multiple control parameters are varied, which is termed multidimensional persistence~\cite{Carlsson2009b}. This problem is a lot more complicated than the single parameter case, due to the absence of simple persistence diagram representations.

We considered examples where point clouds are used to construct undirected graphs and simplicial complex, encoded by matrices with binary elements $\{0, 1\}$, denoting whether a simplex is present or absent. Persistent homology can also be calculated with respect to other fields such as integers modulo 3, describing e.g. directed graphs or simplices, which can be useful for analysing data with twists including points sampled from the surface of M\"obius strips~\cite{ctralie2018ripser}.

\section{Applications of topological data analysis to physics}
\label{sec:physics_applications}

\subsection{Early examples}

Early applications of TDA appearing in physics journals in the 2000s considered examples where the underlying data already has a well-defined shape or graph structure, making the construction of graphs more straightforward. Examples include dynamical systems~\cite{doi:10.1063/1.1705852}, random clouds of spheres~\cite{PhysRevE.74.061107}, random networks~\cite{Horak_2009}, and binary image data~\cite{doi:10.1063/1.2800365,PhysRevLett.107.034503}.

\begin{figure}

\includegraphics[width=\columnwidth]{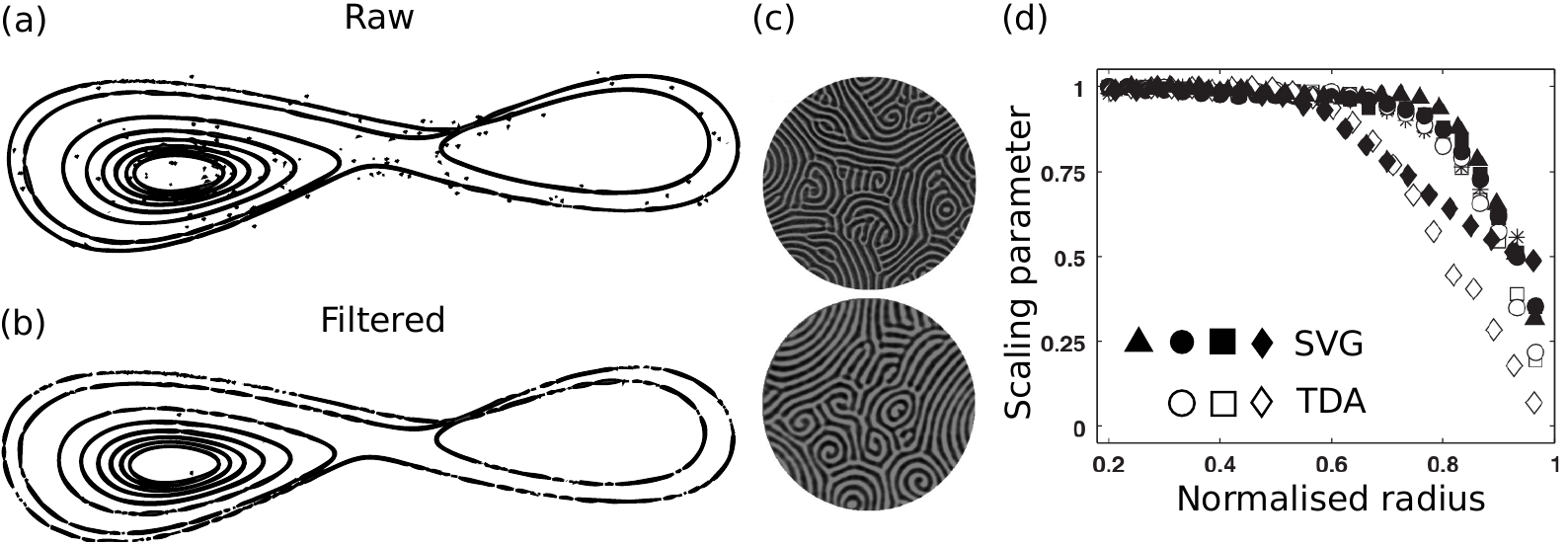}

\caption{(a) Noisy sampling of the chaotic trajectory of the Lorenz attractor and  (b) topology-based filtered data, adapted from Ref.~\cite{doi:10.1063/1.1705852}. (c) Snapshots of a chaotic two-dimensional fluid and (d) estimates of finite-size effects using image eigendecompositions (SVG) and a TDA-based disorder estimator (TDA) showing the two methods give similar results, adapted from Ref.~\cite{PhysRevLett.107.034503}.}

\label{fig:early}

\end{figure}

In the case of over-sampled time-series measurements of dynamical systems, the sampled points will form a single continuous curve in the absence of noise. This fact can be used for topological filtering of certain types of noise, e.g. when a small fraction of the measured points are perturbed, as shown in Fig.~\ref{fig:early}(a). By computing the scale-dependent distribution of zeroth Betti numbers $B_0$ one can separate points belonging to the dynamical trajectory (forming a single big cluster) from noise-perturbed points (each forming a separate cluster), filtering out the latter in Fig.~\ref{fig:early}(b) and improving the accuracy of estimated Lyaponov exponents~\cite{doi:10.1063/1.1705852}.

A second early application was the analysis of convection in two-dimensional fluids under heating~\cite{doi:10.1063/1.2800365,PhysRevLett.107.034503}. There, the fluid separates into distinct hot and cold regions, illustrated in Fig.~\ref{fig:early}(c).  In this case the zeroth and first Betti numbers $B_{0,1}$ were used to characterise the shape of the hot and cold regions. The scaling of the number of distinct microstates (shapes) with the area of the fluid yields an effective dimension of the dynamics that could be computed more efficiently than the conventional approach based on the singular value decomposition of the images' two-point correlation functions. One application of the effective dimension is the detection of boundary effects, shown in Fig.~\ref{fig:early}(d). The strong contrast between hot and cold regions of the images in this case meant that persistent homology was not required; analysis of the graph formed at a single cutoff scale was sufficient.

\subsection{Persistent homology of point clouds and images}

In many applications the construction of well-defined shape from the data is less straightforward, or one may be interested in identifying structures present at different spatial scales. For example, in the case of point cloud data it may be difficult to assign a size or radius to the individual points. In other cases one may want to apply intuition obtained from simple analytically-solvable limits to more realistic systems~\cite{PhysRevLett.126.028102,PhysRevResearch.2.033234}. In situations such as these persistent homology becomes a powerful tool for extracting meaningful shape information from the raw data.

For example, suppose we wish to study the microscopic structure of materials. The raw data naturally takes the positions of the constituent atoms and their sizes. Persistent homology enables studying the multi-scale structure of materials using just the positions of atoms in three-dimensional space (obtained from imaging or simulations) together with the standard Euclidean distance. Refs.~\cite{Nakamura_2015,doi:10.1073/pnas.1520877113} used persistence diagrams computed from molecular dynamics simulations of various materials exhibiting glassy phases to characterise their structure. 

\begin{figure}

\includegraphics[width=\columnwidth]{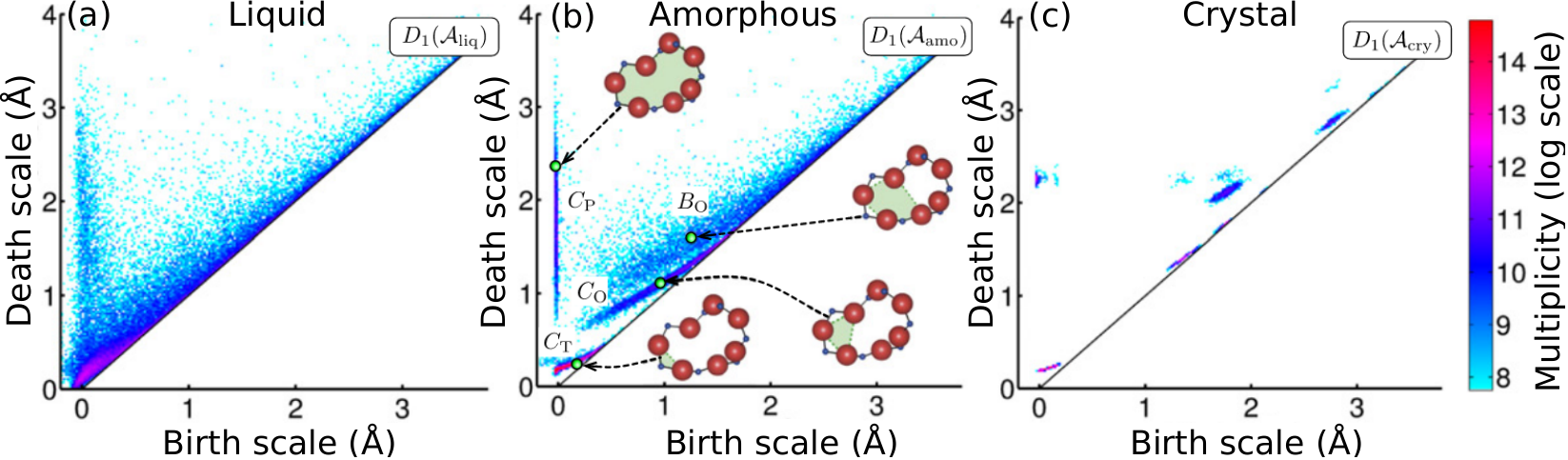}

\caption{Persistence diagrams obtained from molecular dynamics simulations of liquid (a), amorphous (b), and crystalline (c) phases of silica. Point colours indicate the multiplicity (on a logarithmic scale) of one-dimensional features. Insets in (b) illustrate representative cycles corresponding to short- and medium-range order in the amorphous phase. Adapted from Ref.~\cite{doi:10.1073/pnas.1520877113}.}

\label{fig:amorphous}

\end{figure}

Figure~\ref{fig:amorphous} shows examples of persistence diagrams obtained for liquid, glass, and crystalline phases of silica. In the crystalline phase the clustering of feature births and deaths reveal scales corresponding to the bond lengths of the material, i.e. separations between the constituent atoms. Moreover, inspection of the cycles corresponding to persistent features also reveals the nature of the short-range order appearing in the glass phase. 

Subsequent works applied similar techniques to amorphous ices~\cite{Hong_2019}, granular media~\cite{PhysRevE.89.052212,Saadatfar2017}, spin configurations in lattice spin models and gauge theories~\cite{PhysRevResearch.2.043308,PhysRevB.106.085111}, and two-dimensional materials, where the measures obtained using persistent homology can be directly compared with more standard metrics~\cite{doi:10.1063/5.0040393}.

Another application of the point cloud formalism concerns the analysis of time series signals including detection of chaotic dynamics. Already in the 1990s there was interest in applying computational topology to study the shape of the dynamics in phase space, including quantifying the shape of chaotic attractors~\cite{MULDOON19931}. Here the key ingredient is Takens' embedding theorem, which states that a sequence of observations $\phi_t$ taken at regular time intervals $\tau$ can be used to reconstruct the shape of the dynamics by constructing a point cloud of $n$-dimensional vectors $\bs{v}_t = (\phi_t, \phi_{t-\tau}, \phi_{t-2\tau},...)$, provided the embedding dimension $n$ is sufficiently large.

Persistent homology enables the systematic study of dynamics via the shape of the point clouds in the high-dimensional embedding space~\cite{doi:10.1063/1.4949472,doi:10.1063/1.4983840,PhysRevE.99.032209}. For example, period-doubling transitions can be detected via the emergence of new persistent clusters. Successive period-doubling transitions as a system approaches the chaotic regime results in the creation of many clusters, which merge into a single line or volume.

Applying persistent homology to image data enables the study of shapes of images in which there may not be a clear distinction between ``bright'' and ``dark'' regions, or in images where structural information at multiple intensity scales is important. Large point cloud datasets for which a direct persistent homology calculation may be quite time-consuming can alternatively be studied using image filtrations by converting the cloud to a density image~\cite{TEMPELMAN2020132446}. 

Early works on persistent homology of images used Betti numbers to characterise solar magnetic field distributions~\cite{MAKARENKO200798} and force networks in different kinds of compressed granular media, studying the number and connectivity of regions at different scales~\cite{Kondic_2012,PhysRevE.93.062902}. More recently, persistence diagrams obtained from images have been used to study non-Gaussian temperature fluctuations in the cosmic microwave background~\cite{Cole_2018}, the shape of iso-frequency contours in photonic crystals~\cite{doi:10.1063/5.0041084}, many-body dynamics and solitons in Bose-Einstein condensates~\cite{10.21468/SciPostPhys.11.3.060,doi:10.1063/5.0097053}, phase transitions in spin models~\cite{PhysRevB.104.104426,PhysRevE.105.024121}, and order-disorder transitions in nematic liquid crystals~\cite{MembrilloSolis2022} and optical waveguide lattices~\cite{PhysRevB.106.054210}. Recent work aims to better understand how to relate the shape information captured by TDA to physical properties including the permeability of fractured materials~\cite{Suzuki2021}.

\subsection{Finding meaning using abstract distance measures}

So far we have considered examples where we have some intuitive notion of the shape of the underlying data, and the role of TDA has been to study these shapes more systematically. One exciting emerging application of TDA is in studying and discovering structure in complex systems for which simple visualisations (such as images or phase space trajectories) do not exist, including families of high-energy physics models~\cite{Cirafici2016,Cole2019,doi:10.1142/S0217751X20500499}. This typically requires the identification of a suitable distance measure for the data.

For example, in the case of quantum many-body systems measures of entanglement between pairs of subsystems such as the concurrence or entanglement entropy can be used to study the abstract shapes of quantum states and group them into different classes~\cite{di_Pierro_2018,Mengoni2020,PhysRevB.107.115174}. Understanding this entanglement structure may be helpful for judging when approximation techniques such as tensor networks may be used to efficiently simulate the system of interest.

Another important application of abstract distance measures is in the study of condensed matter systems at finite temperatures, where one would like to quantify the ``shape'' of an ensemble of system configurations sampled at a given temperature to detect phase transitions and critical points~\cite{PhysRevE.105.024121,PhysRevE.103.052127,PhysRevB.104.104426,PhysRevResearch.2.043308}. There are various notions of distance that can be applied in this context including the geodesic distance between different spin configurations~\cite{PhysRevE.93.052138} and the quantum distance based on the overlap between eigenfunctions~\cite{doi:10.1063/5.0041084,PhysRevB.105.195115}. Tests of the Anderson, Hubbard, and Potts models suggest TDA may be useful for precisely detecting critical points without requiring computationally-expensive finite size scaling analysis~\cite{PhysRevB.104.235146,Tirelli2022}.

\section{Machine learning for physics using topological data analysis}
\label{sec:TDA_ML}

\subsection{Applications of machine learning to physics}

Machine learning offers powerful data-driven approaches for modelling, characterising, and designing complex physical systems~\cite{RevModPhys.91.045002,MEHTA20191,doi:10.1080/23746149.2020.1797528,https://doi.org/10.48550/arxiv.2204.04198}, including topological materials~\cite{Rodriguez-Nieva2019,PhysRevLett.124.185501,PhysRevLett.124.226401,PhysRevB.102.134213,PhysRevLett.125.127401,PhysRevLett.125.225701,PhysRevLett.130.036601}. Two classes of machine learning approaches attracting interest among physicists are supervised and unsupervised learning algorithms. Supervised learning aims to correctly classify new observations after being trained on a set of labelled examples. Unsupervised learning aims to detect novel features in unlabelled datasets, e.g. by grouping similar observations into clusters or identifying outliers.

Dealing with the deluge of data generated by high energy physics experiments was an early application of large scale machine learning techniques to physics~\cite{Albertsson_2018,RevModPhys.91.045002}. Anomaly-detection techniques are used to identify the small fraction of interesting events to be recorded and processed further. Supervised learning techniques based on human-labelled or computer-generated examples can be used to convert the high-dimensional raw detector data (e.g. particle trajectories and deposited energies) into a signal of interest (e.g. the type of particles generated). Similar techniques are now being adopted in other fields involving high repetition rate experiments, including reconstructing ultrashort optical pulses~\cite{Zahavy:18}, identifying solitons in Bose-Einstein condensates~\cite{Guo_2021}, and optimizing the fidelity of quantum gates~\cite{PRXQuantum.2.040324}. In all these examples, machine learning can be used to perform tasks faster and at a larger scale than conventional approaches. 

The performance of machine learning algorithms is closely tied to the quality and quantity of the input data; the machine learning model needs a sufficiently large set of relevant observations to make accurate predictions. On the one hand, the computational costs of machine learning algorithms can be enormous when they are applied to real-world problems involving large-scale datasets. On the other hand, in many physics problems the amount of available data may be highly constrained (e.g. due to high costs of fabrication, characterization, or computational resources), making approaches compatible with sparse datasets essential.

\subsection{Combining TDA with machine learning}

TDA methods are promising as a means of enhancing the performance of machine learning methods~\cite{10.3389/frai.2021.681108}. Instead of feeding all observables of the system of interest (e.g. entire images or full many-body quantum wavefunctions) into the machine learning algorithm, TDA can identify a smaller set of relevant topological features which can be used as input into a simpler and faster machine learning model. Especially, TDA methods seem naturally suited to studying phenomena such as topological phase transitions or other global structural changes which may be difficult to capture using conventional techniques.

As noted in Sec.~\ref{sec:vectorisation}, a key challenge in combining TDA with machine learning is the question of how best to convert the information encoded into persistence diagrams into a format usable by machine learning algorithms. The two main approaches are distance measure-based and vectorisation.

Refs.~\cite{PhysRevE.103.032207,PhysRevB.104.235146,PhysRevResearch.2.043308,PhysRevB.105.195115,Tirelli2022} have used distance measures of persistence diagrams to compute a distance matrices used as inputs for kernel-based machine learning algorithms for supervised and unsupervised detection of phase transitions in several lattice models including the Ising, XY, and Heisenberg spin models~\cite{PhysRevB.104.235146,PhysRevE.105.024121} and classification of biological time series~\cite{doi:10.1063/1.5125493,PhysRevE.99.032209,PhysRevE.100.032308}. There are many possible metrics to use. In practice the Wasserstein and bottleneck distances~\cite{PhysRevB.104.235146,PhysRevB.105.195115} can be time-consuming to compute. There are faster alternatives  including the sliced Wasserstein distance~\cite{PhysRevResearch.2.043308} and Fisher kernel~\cite{PhysRevE.103.032207}, however they introduce additional hyper-parameters which need to be optimised.

Vectorisation-based approaches can reduce the persistence diagrams into a simpler format that is more easily interpretable, at the expense of losing some of the contained information. For example, Refs.~\cite{PhysRevE.103.052127,doi:10.1063/5.0097053,PhysRevB.106.054210} employed simple summary statistics of the feature lifetimes including their Shannon entropy and norms to verify that persistent homology does indeed detect relevant features that can be used to train machine learning models. Alternatively, Refs.~\cite{PhysRevB.104.104426,PhysRevE.105.024121,ko2022novel} employed persistence images~\cite{adams2017persistence}, which form a discretised representation of persistence diagrams. One word of caution in the use of the persistence images is that their construction involves hyper-parameters that should be optimised to obtain good performance~\cite{10.3389/frai.2021.681108}. 

Once the persistence diagrams have been converted into a format usable by machine learning algorithms, the final step is to choose the specific machine learning model. Several studies have considered supervised classification using logistic regression and support vector machines, which find an optimal separating hyperplane between different data classes~\cite{PhysRevB.104.104426,PhysRevE.105.024121,doi:10.1063/5.0097053}. More recent studies comparing various machine learning models suggest that classification and clustering based on topological features can still be a highly nonlinear problem, making nonlinear machine learning models such as multidimensional scaling, $k$-nearest neighbours, or artificial neural networks a better choice~\cite{PhysRevE.103.052127,PhysRevResearch.2.043308,PhysRevE.105.024121,ko2022novel}. Even when neural network methods are required to obtain an accurate model, the use of TDA-based input features can offer significant reductions in the required width and depth of the networks, making them easier to train~\cite{ko2022novel}.

\subsection{Learning phase transitions using TDA}

The prospect of discovering novel phases of matter motivates studies of machine learning-based approaches for detecting phase transitions. Supervised learning methods can make use of labelled data drawn from known phases or exactly-solvable limits to draw inferences about the location of phase boundaries~\cite{Rem2019,10.21468/SciPostPhys.14.1.005}. On the other hand, unsupervised methods such as manifold learning use an appropriately chosen-similarity measure to compare different samples and group them into different classes, without requiring precise knowledge of the number of distinct phases~\cite{Rodriguez-Nieva2019,PhysRevLett.124.185501,PhysRevLett.124.226401,PhysRevB.102.134213,PhysRevLett.125.127401,PhysRevLett.125.225701,PhysRevLett.130.036601}. 

The reliable detection of phase transitions using machine learning requires use of an appropriate cost function or similarity measure that is sensitive to the transition of interest. For example, topological phase transitions require the bulk band gap of the system to close and re-open, motivating the use of non-local similarity measures invariant under gap-preserving deformations~\cite{PhysRevLett.124.226401} or sensitive to points at which the gap closes~\cite{PhysRevB.102.134213,PhysRevLett.130.036601}. These measures typically-involve hyper-parameters such as the kernel resolution which must be chosen carefully to ensure good accuracy. 

\begin{figure}

\includegraphics[width=0.8\columnwidth]{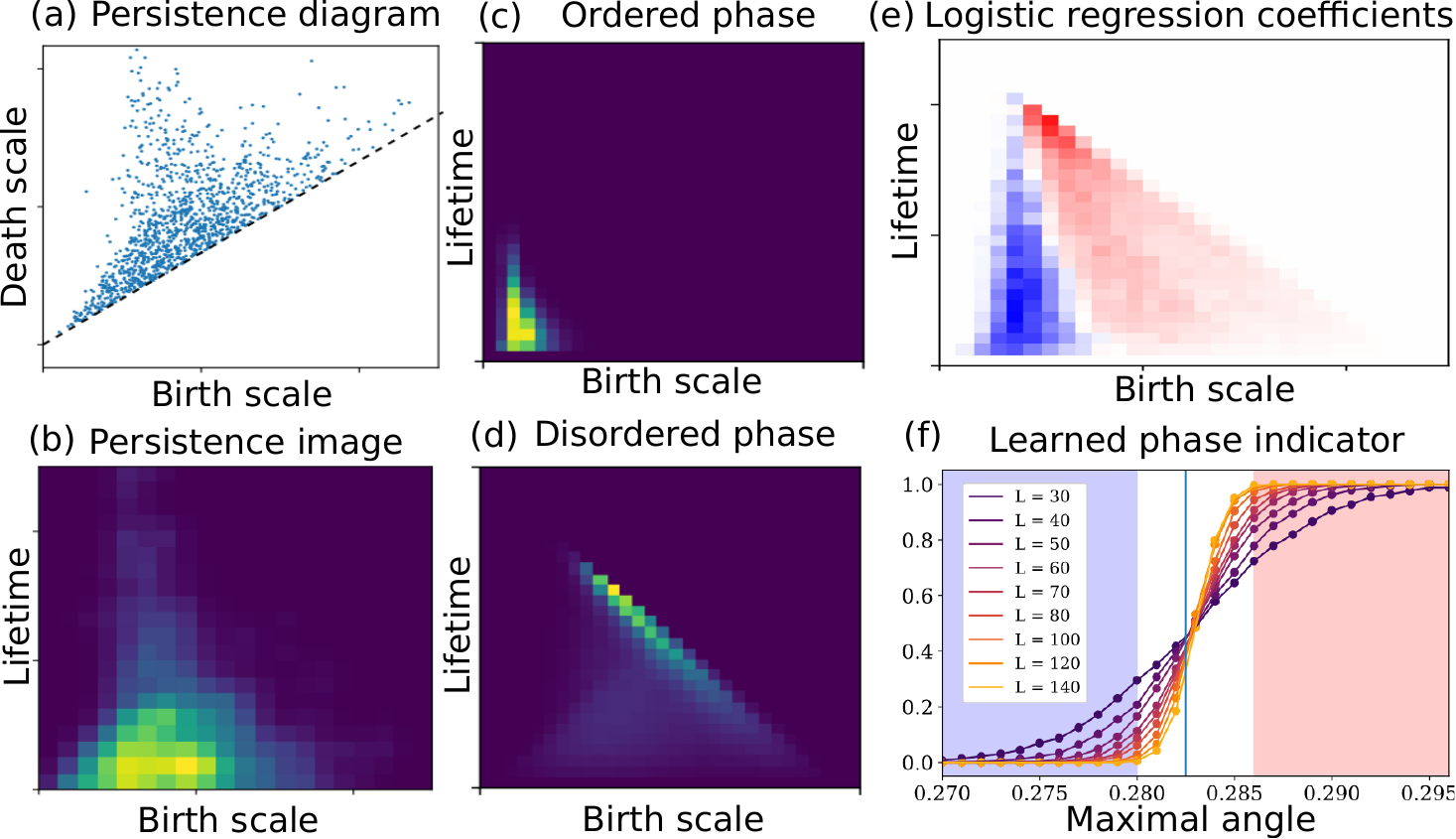}

\caption{TDA-based machine learning of phase transitions in the two-dimensional XY model. (a) Persistence diagram computed using a filtration based on correlations between neighbouring spins. (b) The persistence diagram is vectorised into a persistence landscape. (c,d) Examples of average persistence images obtained from spin configurations sampled from ordered (c) and disordered (d) phases. (e) Coefficients of a trained logistic model; blue and red regions denote parts of the persistence images assigned as indicators to the ordered and disordered phases, respectively. (f) Finite size scaling of the logistic regression prediction used to estimate the phase transition point. Blue and red regions denote ranges used for the model training data. Vertical blue line denotes the expected transition point. Adapted from Ref.~\cite{PhysRevE.105.024121}.}

\label{fig:ML}

\end{figure}

By directly capturing shape information of the system of interest, persistent homology-based methods are able to capture phase transitions using simpler models with fewer hyper-parameters. Figure~\ref{fig:ML} shows an example of a persistent homology-based machine learning pipeline for studying phase transitions in the two-dimensional XY model~\cite{PhysRevE.105.024121}. Persistence diagrams are computed for a given spin configuration based on the relative angle between neighbouring spins. The persistence diagram is then vectorised into a persistence landscape encoding the probability of obtaining features with a given birth scale and lifetime, which can be averaged over spin configurations sampled at a given temperature. The averaged persistence landscapes obtained for various temperatures are then used to train a machine learning model, such as logistic regression, which estimates location of the phase transition using the training data.

\section{Future directions}
\label{sec:future}

\subsection{New techniques for topological data analysis}

An area of active research among physicists is the application of TDA tools to analyse the structure of more complex systems including flow networks involving directed links~\cite{PhysRevE.105.044311} and time-evolving networks~\cite{PhysRevE.100.032308}. One approach used in recent studies that is compatible with standard persistent homology tools is to convert the directed network into a regular point cloud using a diffusion map, which constructs edges between a pair of vertices $(i,j)$ by computing the probability of diffusion between $i$ and $j$. It will be interesting to explore alternate approaches that can work directly with unidirectional or time-evolving systems without requiring diffusion maps, such as zigzag persistence~\cite{https://doi.org/10.48550/arxiv.2205.11338}.

The metrics used for quantifying differences between persistence diagrams have applications beyond persistent homology. For example, Ref.~\cite{PhysRevLett.126.048101} used the Wasserstein distance to compare different local neighbourhood structures of disordered media, based on the intuition that it encodes the energy cost required to transform one configuration to another. The advantage of such a topological metric compared to more conventional measures including the Kullback-Leibler divergence is that the former is better at distinguishing weakly-overlapping distributions. Are there other examples where such metrics can be linked to physical observables?

\subsection{Quantum topological data analysis}

All of the examples considered in the physics literature relate to the study of low-dimensional topological features using TDA, largely because higher dimensional features are both harder to interpret and become extremely time-consuming to compute for large datasets, due to exponential scaling. The advent of more efficient quantum algorithms for TDA including computation of Betti numbers and persistence diagrams is anticipated to enable the study of higher-dimensional topological features of complex datasets.

The first quantum algorithm for TDA was proposed by Lloyd et al. in 2016~\cite{Lloyd2016}. Their algorithm exhibited an exponential speedup for calculating Betti numbers by using quantum phase estimation to efficiently construct combinatorial Laplacians of simplicial complexes and identify cycles by computing their zero modes. This proposal was followed in 2018 by a small-scale few-qubit proof-of-concept quantum optics experiment~\cite{Huang:18}.

Subsequent studies have started to address limitations of the first quantum TDA algorithm~\cite{https://doi.org/10.48550/arxiv.1911.10781} by proposing more efficient variants~\cite{https://doi.org/10.48550/arxiv.2209.12887,https://doi.org/10.48550/arxiv.2209.13581,https://doi.org/10.48550/arxiv.2209.10596} as well as quantum algorithms for computing persistent Betti numbers~\cite{https://doi.org/10.48550/arxiv.2202.12965,Hayakawa2022quantumalgorithm} and the Wasserstein distance~\cite{https://doi.org/10.48550/arxiv.1809.06433}. While most of these algorithms are designed for future fault-tolerant quantum computers, there is also the potential for near-term quantum speedups using shallow quantum circuits with depth linear in the number of input data points~\cite{https://doi.org/10.48550/arxiv.2108.02811,https://doi.org/10.48550/arxiv.2209.09371}, exploiting the efficient implementation of the boundary operator using entangled quantum states~\cite{cade2021complexity,PhysRevA.106.022407,https://doi.org/10.48550/arxiv.2202.00054}. While the prospect of an exponential speed up compared to the best classical TDA algorithms is entrancing, whether and when such a large speed for practical problems will be achieved is under debate~\cite{Gyurik2022towardsquantum,https://doi.org/10.48550/arxiv.2209.13581,https://doi.org/10.48550/arxiv.2209.09371,https://doi.org/10.48550/arxiv.2209.12887,https://doi.org/10.48550/arxiv.2209.14286}, especially with new and improved classical algorithms still being developed~\cite{https://doi.org/10.48550/arxiv.2211.09618}.

\subsection{Learning physics versus machine learning physics}

One challenge encountered by existing literature applying TDA to physics problems is that the techniques are unfamiliar to the physics audience. Many of the original TDA articles in which techniques were first introduced are highly theoretical and mathematically-rigorous, thus articles in physics journals require long introductions explaining the approaches used to this non-specialist audience. This can lead to a focus on the technical calculation details while perhaps obscuring the bigger picture.

For instance, many articles include examples of persistence diagrams computed from the system of interest in order to illustrate qualitative differences between different phases or states. However, the persistence diagram is itself not easily interpretable, requiring knowledge of the form of the input data and filtration used. For this reason, many studies then apply machine learning techniques to extract quantitative predictions from the information contained in the persistence diagrams.

On the other hand, as physicists we would prefer to make sense of the system ourselves, rather than delegate understanding to a machine learning algorithm. What is of interest to us are which topological features are meaningful, and what they look like. The approach used in Ref.~\cite{doi:10.1073/pnas.1520877113}, where representative cycles of the persistent features are included as insets in the persistence diagrams [reproduced here in Fig.~\ref{fig:amorphous}(b)], is one way their meaning can be made more explicit. Still, selecting appropriate features can be challenging when there is no clear boundary between the signal and the noise. It will be interesting to explore other TDA-based techniques for dimensionality reduction and compactly conveying the significant features of high-dimensional physical systems to non-specialists in TDA.

\section{Conclusion}
\label{sec:conclusion}

In summary, we have given an overview of emerging physics applications of topological data analysis, focusing on persistent homology. The take-home message is that TDA can be used to compress complex datasets into their essential (topological) features which can be used as input to simpler machine learning models compared to widely-used and computationally-expensive artificial neural networks. Nevertheless, as topological data analysis is relatively new it is still largely employed on an ad-hoc basis and further work is needed to establish a standard set of methods that non-specialists can trust~\cite{10.3389/frai.2021.681108}.

Topological data analysis has already been fruitfully applied to other areas of research including image analysis and medical science, enabling the extraction of useful insights from complicated hard-to-visualise datasets. For example, TDA-based methods have outperformed other more popular machine learning approaches for complicated problems such as predicting biomolecule binding efficiencies~\cite{doi:10.1126/sciadv.abc5329}. We hope that the techniques discussed here and in other recent reviews aimed at the physics audience~\cite{Carlsson2020, https://doi.org/10.48550/arxiv.1904.11044} will not merely provide a transient fashionable alternative to more standard methods of data analysis used by physicists, but will form a new set of long-lasting tools enabling a better understanding of complex physical systems from classical to quantum.

\section*{Disclosure statement}

We declare no potential conflicts of interest.

\section*{Funding}

This research is supported by the National Research Foundation, Singapore and A*STAR under its CQT Bridging Grant and Quantum Engineering Programme
NRF2021-QEP2-02-P02, A*STAR (\#21709) and by EU HORIZON - Project 101080085 — QCFD.

\bibliography{TDA_refs}

\end{document}